\documentclass[12pt]{iopart}

\expandafter\let\csname equation*\endcsname\relax
\expandafter\let\csname endequation*\endcsname\relax
\usepackage{amsmath}
\usepackage{graphicx}
\usepackage{epstopdf}
\usepackage{multirow}
\usepackage{subfigure}

\usepackage{iopams,cite}
\input xy
\xyoption{all}
\usepackage{amsthm}
\usepackage{amsfonts,amsmath}
\usepackage{lscape}

\theoremstyle{definition}

\theoremstyle{remark}

\begin{document}
\title{Energy-level crossings and number-parity effects in a bosonic tunneling model}
\author{
Davids Agboola, Phillip S. Isaac, and Jon Links}
\address{Centre for Mathematical Physics, School of Mathematics and Physics, \\
The University of Queensland 4072,
 Australia}

\eads{\mailto{d.agboola@maths.uq.edu.au, psi@maths.uq.edu.au, jrl@maths.uq.edu.au}}

\date{}


\begin{abstract}
\noindent  
An exactly solved bosonic tunneling model is studied along a line of the
coupling parameter space, which includes a quantum
phase boundary line. The entire energy spectrum is computed
analytically, and found to exhibit multiple energy level crossings in a
region of the coupling parameter space. Several key properties of the model are
discussed, which exhibit a clear dependence on whether the particle number is even or odd.
\end{abstract}


\section{Introduction}

The symmetric two-site Bose--Hubbard model has been studied widely for some time \cite{Mil97,Cir98,Leg01,Kohl02,Zhou03,Pan05}. The Hamiltonian reads
\begin{align}
H=\frac{k}{8}(N_1-N_2)^2 -\frac{J}{2}(b_1^\dagger b_2 + b_2^\dagger b_1)
\label{aaham}
\end{align}\
where 
\begin{equation*}
[{b}_{i},{b}_{j}^{\dagger}]=\delta_{ij}{\mathcal{I}},\quad [{b}_{i},\,{b}_{j}]=[{b}_{i}^{\dagger},\,{b}_{j}^{\dagger}]=0,
\end{equation*}
 for $i,j=1,2$. Above   $\mathcal{I}$ denotes the identity operator, and
 ${N}_{j}={b}_{j}^{\dagger}{b}_{j}$. Setting ${N}={N}_{1}+{N}_{2}$, it can be
 verified that $[H,\,{N}]=0$. The model has a simple interpretation through two
 terms describing particle interactions with coupling $k$, and a tunneling
 process between two wells with interaction strength $J$. Without loss of
 generality we take $J\geq 0$. Though it is simple, the Hamiltonian has been
 successfully used as a model for experimentally realised tunneling phenomena
 \cite{albiez05}. 

Several studies  have identified a quantum phase transition in the attractive
regime $k<0$, using a variety of approaches including  semiclassical methods \cite{Zib10,Sim12},  mean-field approximation \cite{Gra14}, entanglement \cite{Perez10,rubeni12,Buo12}, fidelity
\cite{rubeni12,Buo12}, fragmentation \cite{Polls10,Sak14}, NMR simulations \cite{izi15}, and exact results using Bethe Ansatz methods \cite{rubeni12,lm15}.  One way to
characterise the two phases is through the energy gap between the ground state
and the first excited state. Setting $k=0$ in (\ref{aaham}) it is not difficult
to check that the ground-state energy is $-JN/2$, and the gap to the first
excited state is $J$. At the other extreme when $J=0$ and $k<0$, the ground
state is two-fold degenerate, so the gap is zero. The transition between these
extremes is abrupt. Setting $\lambda=({kN})/(2J)$
the transition takes place at $\lambda=-1$ \cite{rubeni12}.  

In recent times a generalised version of (\ref{aaham}) has been studied which
includes a second-order tunneling process \cite{llll09, cf12, rlif17}.
The extended Hamiltonian is 
\begin{align}
H=\frac{k}{8}(N_1-N_2)^2 -\frac{J}{2}(b_1^\dagger b_2 + b_2^\dagger b_1)
-\frac{\Omega}{2}((b_1^\dagger)^2 b^2_2 + (b_2^\dagger)^2 b^2_1)
\label{ham}
\end{align}
where the coefficient $\Omega$ is the coupling for second-order tunneling. The
inclusion of such a term can be justified on physical grounds, but it often
neglected because the coupling $\Omega$ is much weaker than $k$ and $J$
\cite{ab06,go07}. Nonetheless, the model has been employed \cite{llll09} to account for the
observation of second-order tunneling in the low-particle number limit \cite{folling07}.
From the mathematical perspective, (\ref{ham}) offers a richer structure than
(\ref{aaham}). Analyses of bifurcations of fixed points in the classical limit
show there are three expected phases, which will be referred to as {\it
Josephson}, {\it self-trapping}, and {\it phase-locking} \cite{cf12, rlif17}.
Multiple energy-level crossings were found in the phase-locking phase through
the studies of \cite{rlif17}. Such crossings are a new feature not found in the studies of the
Hamiltonian (\ref{aaham}).

The main objective of this work is to investigate the boundary between the phase-locking and
self-trapping phases. Energy-level crossings are also found to occur on this
boundary, and they can be precisely identified. The energy levels can be
computed analytically.  The character of the set of energy levels is dependent on
whether the particle number is even or odd. We will study some of the
consequences of this finding, which may have implications for few-body
bosonic systems. The results complement those for few-body fermions systems, that have attracted recent attention 
\cite{zwmblj13,ss16}.

In Sect. 2 we begin by establishing that the phase-locking and self-trapping
phases exhibit a duality. 
The boundary between them is a self-dual line with an enhanced symmetry. 
In Sect. 3 we recall a Bethe Ansatz equations for the model, which are easily solved on the self-dual line. 
This solution is used in Sect. 4 to examine the nature of the ground-state energy
gap, and in Sect. 5 a supersymmetric structure within the model is unveiled. 
Number-parity effects in the computation of dynamical expectation values are investigated in Sect. 6, and 
concluding remarks are given in Sect. 7.


\section{Duality}

Set $\gamma=\Omega N/J$, and recall $\lambda=(kN)/(2J)$.
The boundary lines between the three phases, which are identified through bifurcation analysis, are (see also Fig. 6 in \cite{rlif17}) 
\begin{itemize}
\item[$\bullet$] Self-Trapping/Josephson: $\gamma+\lambda=-1$ for $\lambda\leq -1/2$;
\item[$\bullet$] Phase-Locking/Self-Trapping: $\gamma=\lambda$ for $\lambda\leq -1/2$;
\item[$\bullet$] Josephson/Phase-Locking: $\gamma=-1/2$ for $\lambda\geq -1/2$.
\end{itemize}
The three boundaries meet at the triple point $(\lambda,\gamma)=(-1/2,-1/2)$. 

To reveal the duality between the  phase-locking and self-trapping phases, introduce  the $su(2)$ realisation
\begin{align}
S^+=b_1^\dagger b_2,\,\quad S^-=b_2^\dagger b_1,\,\quad S^z=(N_1-N_2)/2
\label{su2real}
\end{align} 
satisfying the relations 
\begin{align}
 [S^z, S^\pm]=\pm S^\pm,\qquad [S^+,S^-]=2S^z,
\label{su2}
\end{align} 
for which the Casimir invariant $C=2(S^z)^2+S^+S^- + S^-S^+$ has eigenvalue $\Lambda=N(N+2)/2 $.
In terms of this realisation, the Hamiltonian is expressed as
\begin{align}
H&=\frac{k}{2}(S^z)^2-JS^x-\frac{\Omega}{2}((S^+)^2+(S^-)^2).
\label{lmg}
\end{align}
This Hamiltonian will now be transformed by a composition of three unitary operators:
\begin{align*}
&T: S^x \mapsto -S^z,\,S^y \mapsto S^y,\,S^z \mapsto S^x , \\
&R: S^x \mapsto -S^y,\, S^y \rightarrow S^x,\, S^z \rightarrow S^z,  \\
&U=T^{-1}\circ R \circ T
\end{align*}
where $S^x=(S^++S^-)/2, \, S^y=(S^+-S^-)/(2i) $ 
such that $U^4={\rm id}$.
It is found that
\begin{align*}
U(H)
&=\frac{1}{4}\left({6\Omega}-{k}\right)(S^z)^2-JS^x-\frac{1}{8}\left({k}+2{\Omega} \right)
\left((S^+)^2+(S^-)^2\right) 
+\frac{1}{8}\left({k}-  2\Omega  \right)C, 
\end{align*}
and $U^2(H)=H$. It is easily checked that, up to the inclusion of an
$N$-dependent term, $U$ maps Hamiltonians between the phase-locking and
self-trapping phases, while Hamiltonians in the Josephson phase are mapped back
to the Josephson phase under the action of $U$. This shows that there is a 1-1
correspondence between the energy spectra in phase-locking and self-trapping
phases.
Hamiltonians on the line $\gamma=\lambda$, or equivalently $\Omega=k/2$, are
invariant under the action of $U$. Along this line, which   includes the
boundary between the phase-locking and self-trapping phases, analytic
expressions for the entire energy spectrum can be obtained, as we describe
below.

\section{Exact solution}
The Bethe Ansatz solution derived in \cite{rlif17} gives the energy eigenvalues and eigenvectors as 
\begin{align*}
E&= \frac{kN^2}{8}-\frac{J}{2}\sum_{j=1}^N u_j
	-\frac{\Omega}{2}\sum_{j=1}^N\sum_{k\neq j}^Nu_j u_k, \\
|\Psi\rangle&=\prod_{j=1}^N(b^\dagger_1 +u_j b^\dagger_2)|0\rangle.  
\end{align*}
Here, the parameters $\{u_j:j=1,...,N\}$ satisfy the Bethe Ansatz Equations (BAE)
\begin{align}
(J(1-u_j^2)-k(N-1)u_j+2\Omega(N-1)u_j^3 ) Q'(u_j)=   (\Omega(1+u_j^4)-ku_j^2)Q''(u_j)
\label{bae}
\end{align}
where 
\begin{align}
Q(x)=\prod_{j=1}^N(x-u_j).
\label{que}
\end{align}
Note that the form (\ref{bae}) is different to the BAE presented in \cite{rlif17}, which reads
\begin{align}
\frac{J(1-u_j^2)-k(N-1)u_j+2\Omega(N-1)u_j^3}{ku_j^2-\Omega(1+u_j^4)}&=\sum_{k\neq j}^N\frac{2}{u_k-u_j}.
\label{baealt}
\end{align}
Eqs. (\ref{bae}) and (\ref{baealt}) are equivalent whenever there are no root multiplicities in (\ref{que}) . The more general form 
(\ref{bae}), which accommodates root multiplicities, will be required for the analysis below.

Hereafter set $\Omega=k/2$, which is the self-dual line identified in the
previous section. For this constraint the BAE (\ref{bae}) are solved
with the choice $u^2_j=1$ for all $j=1,\ldots,N$. There are $N+1$ solutions
where $q$ of the roots are chosen to taken the value $-1$, while the remaining
$N-q$ are chosen to take the value $1$.
This gives a complete set of (normalised) eigenstates
\begin{align}
|N,q\rangle&=\frac{1}{\sqrt{2^{N}q!(N-q)!}}(b_1^\dagger+b_2^\dagger)^{N-q}(b_1^\dagger -b_2^\dagger)^{q}|0\rangle 
\label{nstates}
\end{align}
with the corresponding energies
\begin{align}
E(N,q)&=\frac{J}{2}(2q-N)+\frac{k}{8}(8q(N-q)+2N-N^2) .  
\label{nrg}
\end{align}

While the structure of the states and spectrum through
(\ref{nstates},\ref{nrg}) is very simple, we can see that the system is
non-trivial from the following analysis. Fixing $J$ in (\ref{nrg}), by setting
$E(N,q)=E(N,q')$, we see
that all energy levels corresponding to labels
$q,q'\in\left\{0,1,\ldots,\left\lfloor
\frac{N}{2}\right\rfloor\right\}$ will cross at values 
\begin{equation}
k = \frac{J}{q+q'-N}.
\label{cross}
\end{equation}
Moreover, energy levels corresponding to $q$ and $q'=q+1$ cross when 
\begin{equation}
k=\frac{J}{2q+1-N},
	\label{kgroundstate}
\end{equation}
which decreases as $q$ increases. This then implies that for all
${k>J/(1-N}),$ the label $q=0$ corresponds to the ground
state, noting that for these values of $k$ there are no further energy level crossings for this state. By a similar
argument, for all $k<J/(2\left\lfloor \frac{N}{2}\right\rfloor - 1 - N),$ the label 
$q=\left\lfloor\frac{N}{2}\right\rfloor$ corresponds to the ground state. For labels
$q=1,2,\ldots,\left\lfloor\frac{N}{2}\right\rfloor-1$, the ground
state occurs when 
\begin{equation}
\frac{J}{2q+1-N} < k < \frac{J}{2q-1-N}.
	\label{qgsk}
\end{equation}
This is easily seen using standard calculus techniques. In other words, all the
ground state energy level crossings occur from the lowest value 
$k=J/(2\left\lfloor\frac{N}{2}\right\rfloor - 1 - N)$ up to $k=J/(1-N).$

The level crossings predicted by our analysis can be seen in 
Fig. 1. From the diagram we can see that when $k=0$ the energies are equally spaced, e.g. see Fig. 1(a). 
For negative $k$, as $|k|$
increases, a sequence of level crossings occurs, e.g. see Fig. 1(b). Also for
negative $k$, for sufficiently large
$|k|$ the energies form a system of bands. When $N$ is odd the number of
energy levels is even, and the energy level bands occur in pairs $|N,q\rangle$
and $|N,N-q\rangle$, each with separation $J(N-2q)$. When $N$ is even, however, 
the number of levels is odd, and there is a single unpaired state, e.g. see Fig. 1(c).
This points towards a prospect for number-parity effects, which will be
explored below. 
 


\begin{figure}[h]
\subfigure[]{\includegraphics[scale=0.29]{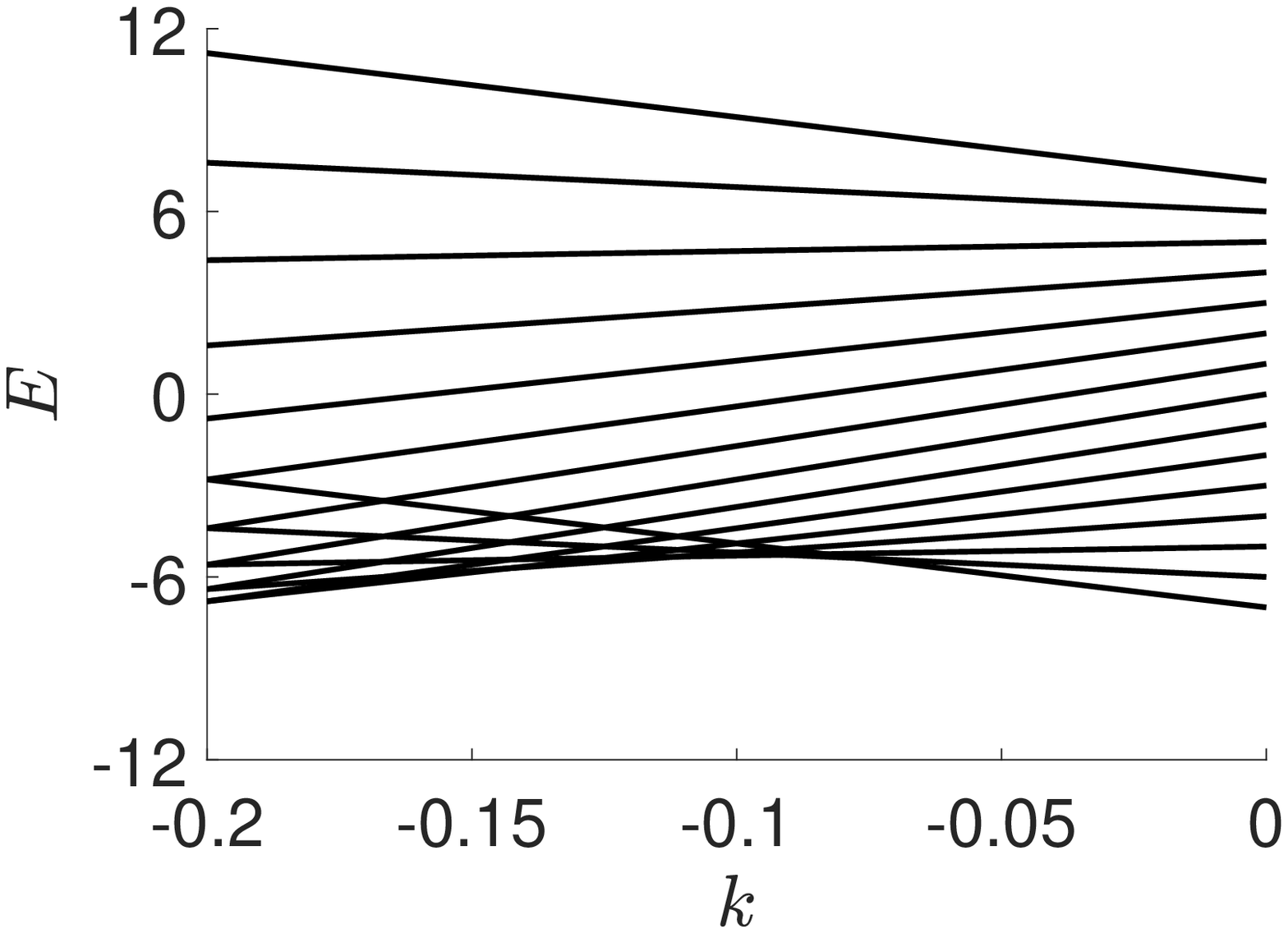}} 
\subfigure[]{\includegraphics[scale=0.29]{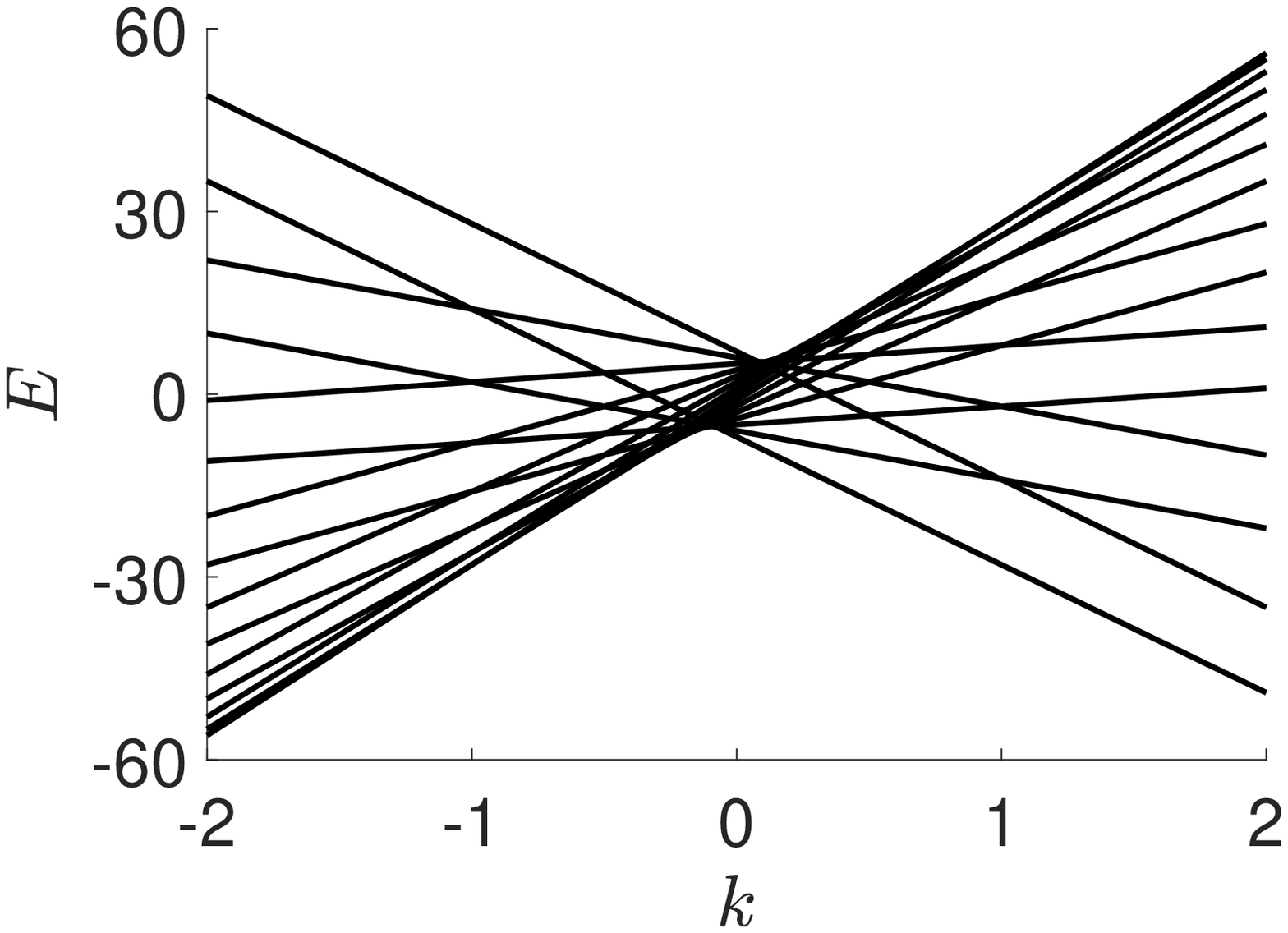}} 
\subfigure[]{\includegraphics[scale=0.29]{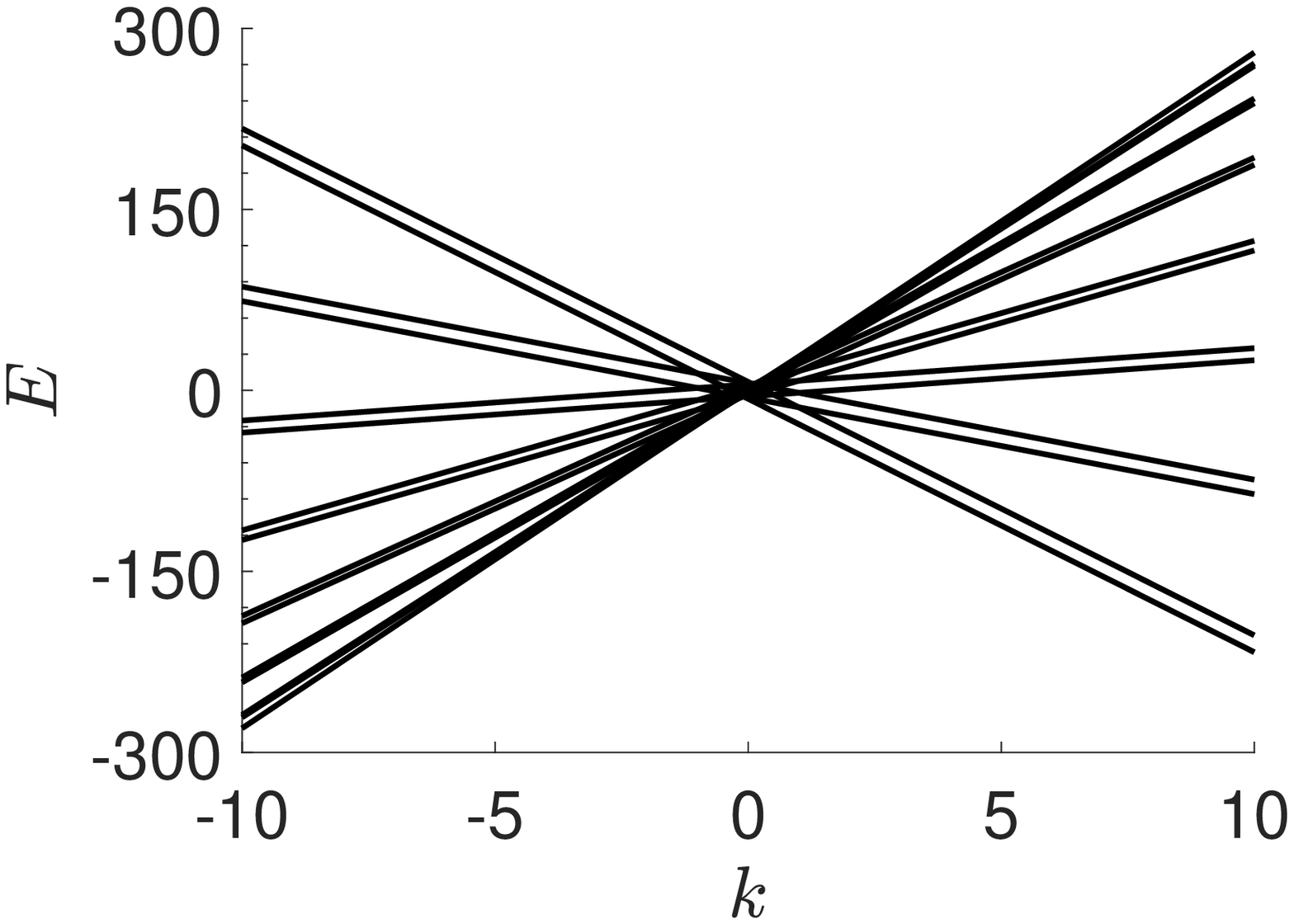}}
\begin{caption}
{Energy levels as a function of $k$ for $J=1$, $\Omega=k/2$ and $N=14$. The figures depict different orders of magnitude for the interval of  $k$.} 
\end{caption}
\label{nrgN6figs}
\end{figure}

We also remark that from (\ref{nstates}), it is straightforward to calculate certain correlation functions. For example, for each $q=0,1,...,N$
\begin{align*}
\langle(N_1-N_2)^2\rangle&= N+2q(N-q),      \\
\langle b_1^\dagger b_2+b_2^\dagger b_1\rangle&=N-2q,  \\
\langle (b_1^\dagger)^2b_2^2+(b_2^\dagger)^2b_1^2\rangle&=\frac{N^2}{2}-\frac{N}{2}-3q(N-q).
\end{align*}

\subsection{Continuum approximation}

One straightforward approach to analyse the system is to introduce the variable  
$l=q/N$, $0\leq l \leq 1$,  and treat this as varying continuously. This approximation is expected to be a valid in the limit of large 
$N$. To leading order in $N$ (\ref{nrg}) becomes 
\begin{align*}
	E&\approx\frac{JN}{2}(2l-1)+4\lambda(l(1-l)) -\frac{JN\lambda}{4} .  
\end{align*}
For $\lambda \geq -1/2$ the minimum value of energy occurs at $l=0$, while for
$\lambda\leq -1/2$, the minimum occurs at 
\begin{align}
l=\frac{1}{2}+\frac{1}{4\lambda}.
\label{ell}
\end{align}
We note that (\ref{ell}) is consistent with (\ref{kgroundstate}) in the
large $N$ limit. 
The following expression are then found for the ground-state energy and correlations:
\begin{align*}
\frac{E_0}{N}&\approx\begin{cases}
	-\frac{J}{4}(2+\lambda),  &     \lambda\geq -\frac{1}{2}, \\
	\frac{J}{8}( 2\lambda + \frac{1}{\lambda}),  &     \lambda\leq -\frac{1}{2}, 
\end{cases} \\
\frac{\langle (N_1 -N_2)^2\rangle}{N^2}&\approx\begin{cases}
0,  &     \lambda\geq -\frac{1}{2}, \\
\frac{1}{2}-\frac{1}{8\lambda^2}, \phantom{a)} &     \lambda\leq -\frac{1}{2}, 
\end{cases} \\
\frac{\langle b_1^\dagger b_2+b_2^\dagger b_1\rangle}{N}&\approx\begin{cases}
1,  &     \lambda\geq -\frac{1}{2}, \\
-\frac{1}{2\lambda},  \phantom{aaaa)} &     \lambda\leq -\frac{1}{2}, 
\end{cases} \\
\frac{\langle (b_1^\dagger)^2b_2^2+(b_2^\dagger)^2b_1^2\rangle}{N^2}&\approx\begin{cases}
\frac{1}{2},  &     \lambda\geq -\frac{1}{2}, \\
-\frac{1}{4}+\frac{3}{16\lambda^2},  \phantom{}&     \lambda\leq -\frac{1}{2}, 
\end{cases} 
\end{align*}  in this subsection 
The fact that the value of $l$  as given by (\ref{ell}) is a function of
$\lambda$ is a reflection of the level crossings. It is also has the effective
a treating the system a being gapless when $\lambda\leq -1/2$. While this is
correct in some sense, the above treatment does not capture the full physical
properties of the model.

\section{Ground-state energy gap}
Define the ground-state energy gap $\Delta$ to be the difference between the first excited-state energy and the ground-state energy. 
From previous discussion, we know that for fixed $J$ and
$q=1,2,\ldots,\left\lfloor\frac{N}{2}\right\rfloor-1$, the energy level $|N,q\rangle$ is the
ground state for values of $k$ given by (\ref{qgsk}). In the following, we only consider
this range of $k$ and $q$ values, so that $|N,q\rangle$ is the ground state. 
In this case, it is straightforward to show that $E(N,q+1)=E(N,q-1)$ occurs
when 
\begin{align}
k =k_P &= \frac{J}{2q-N}.
	\label{kcrit}
\end{align}
The difference in the energy corresponding to $k_P$ and the ground
state is found to be $J/(N-2q)$. Also, the difference in energy between the
ground state $|N,q\rangle$ and the state $|N,N-q\rangle$ is $J(N-2q)$ which is
greater than $J/(N-2q)$ for the given values of $q$. It
follows that peaks in the gap must occur at the values $k_P$ given in
(\ref{kcrit}), corresponding to the crossing of $|N,q-1\rangle$ and
$|N,q+1\rangle$.

\begin{figure}[h]
\center{
\subfigure[]{\includegraphics[scale=0.27]{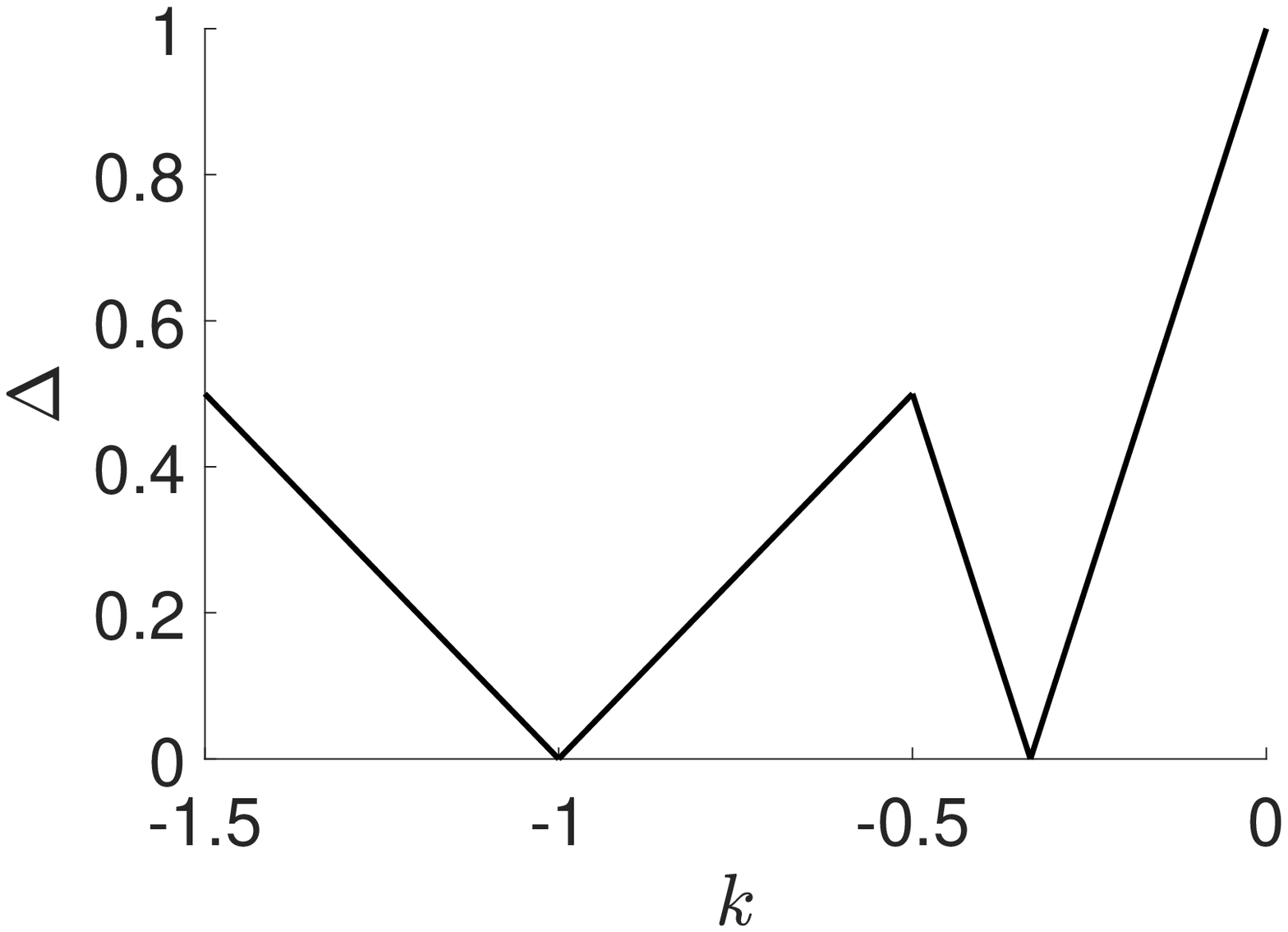}} 
\subfigure[]{\includegraphics[scale=0.27]{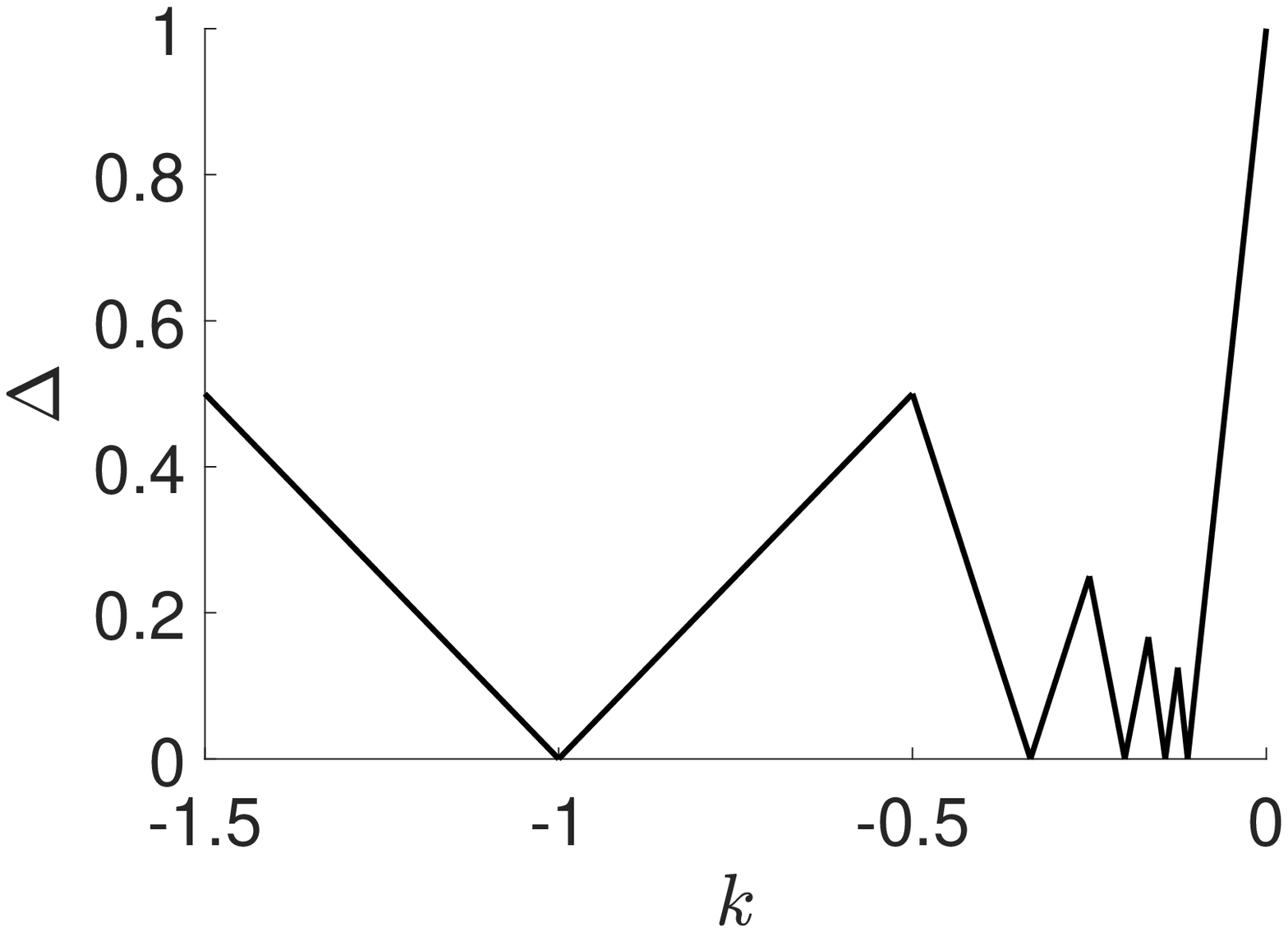}} 
\subfigure[]{\includegraphics[scale=0.27]{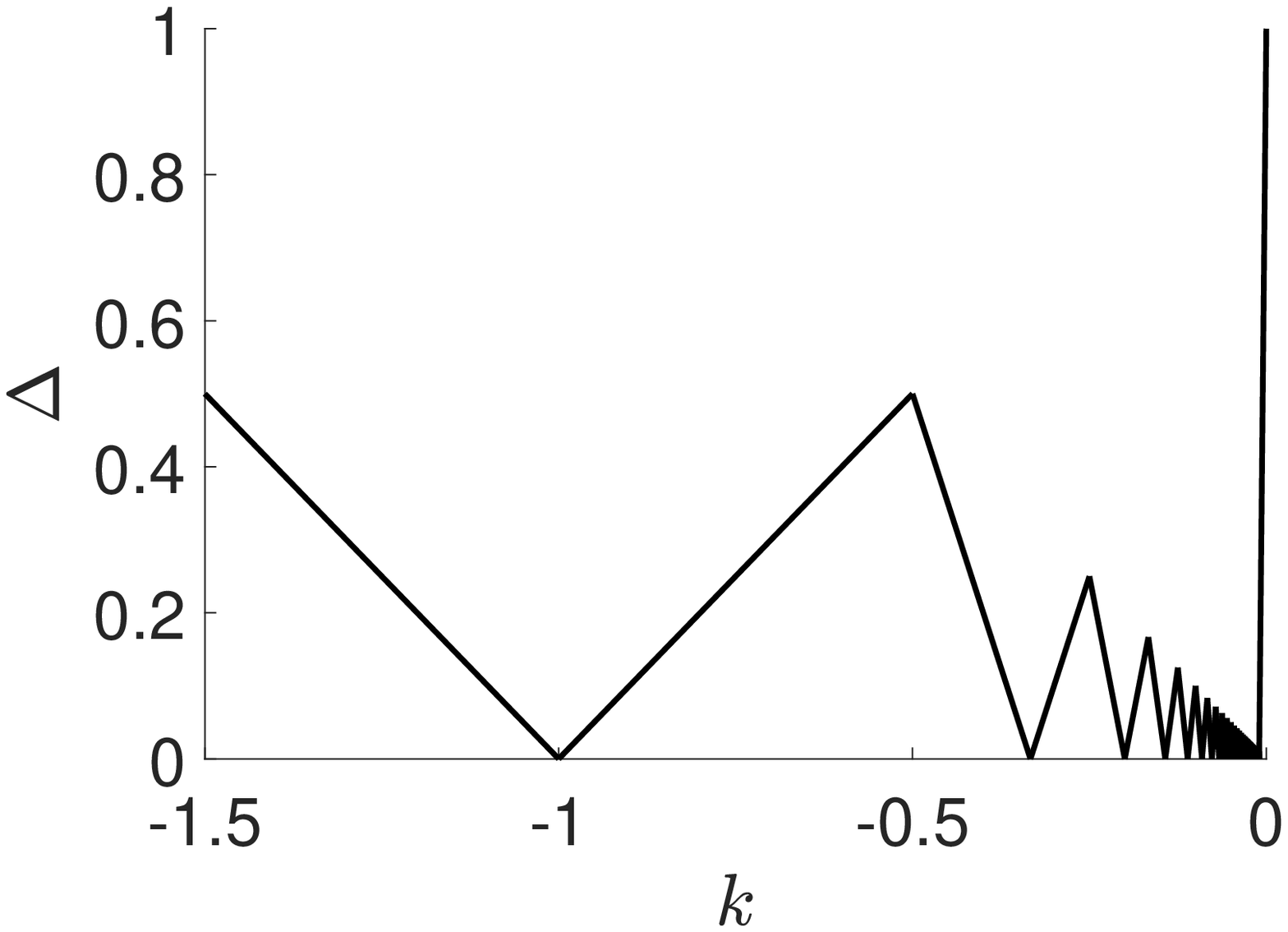}}
}
\begin{caption}
{Ground-state energy gap $\Delta$ as a function of $k$ with $J=1$, and $\Omega=k/2$. (a) $N=4$, (b) $N=10$, (c) $N=100$. }
\end{caption}
\end{figure}

\begin{figure}[h]
\center{
\subfigure[]{\includegraphics[scale=0.27]{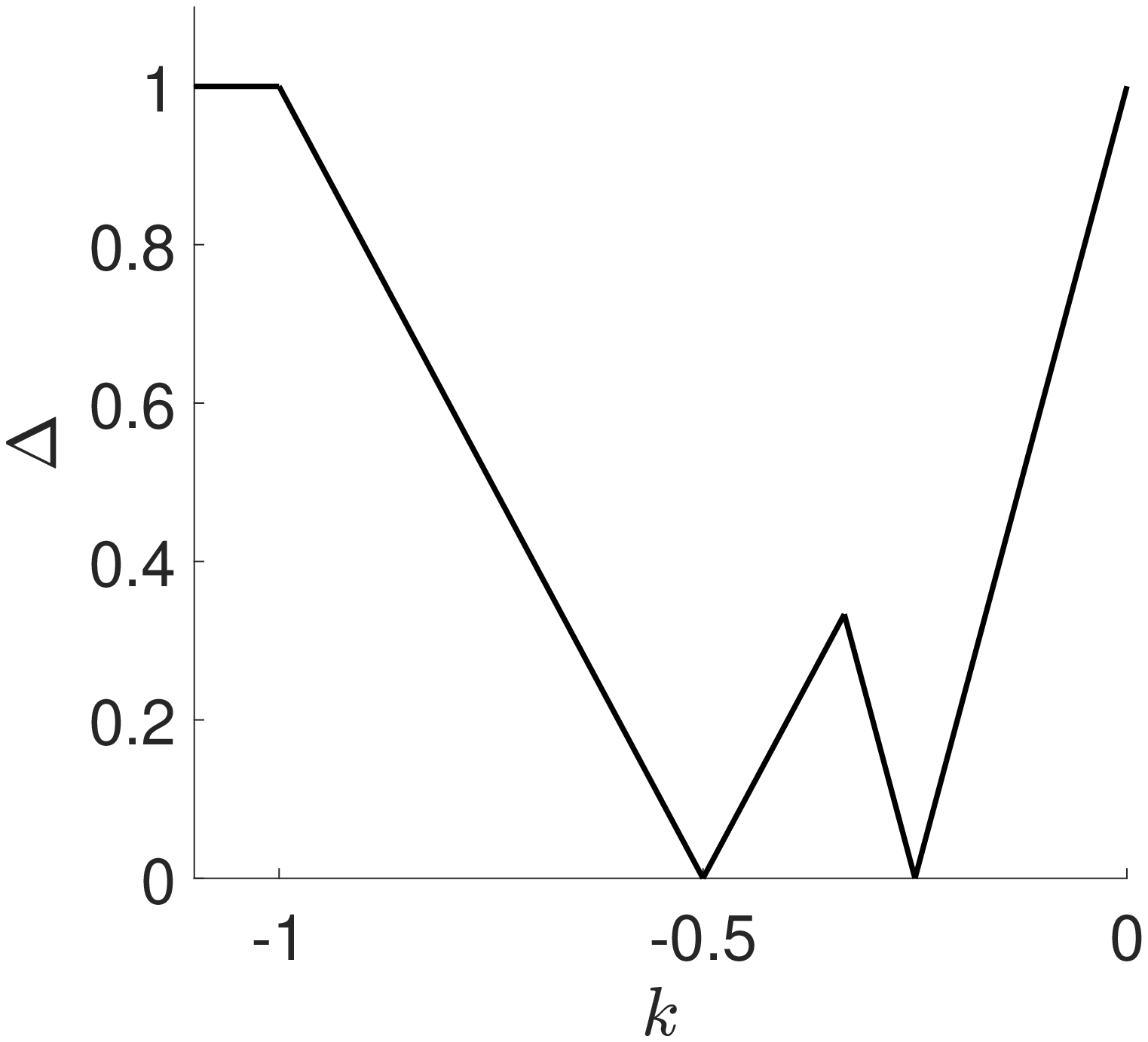}} 
\subfigure[]{\includegraphics[scale=0.27]{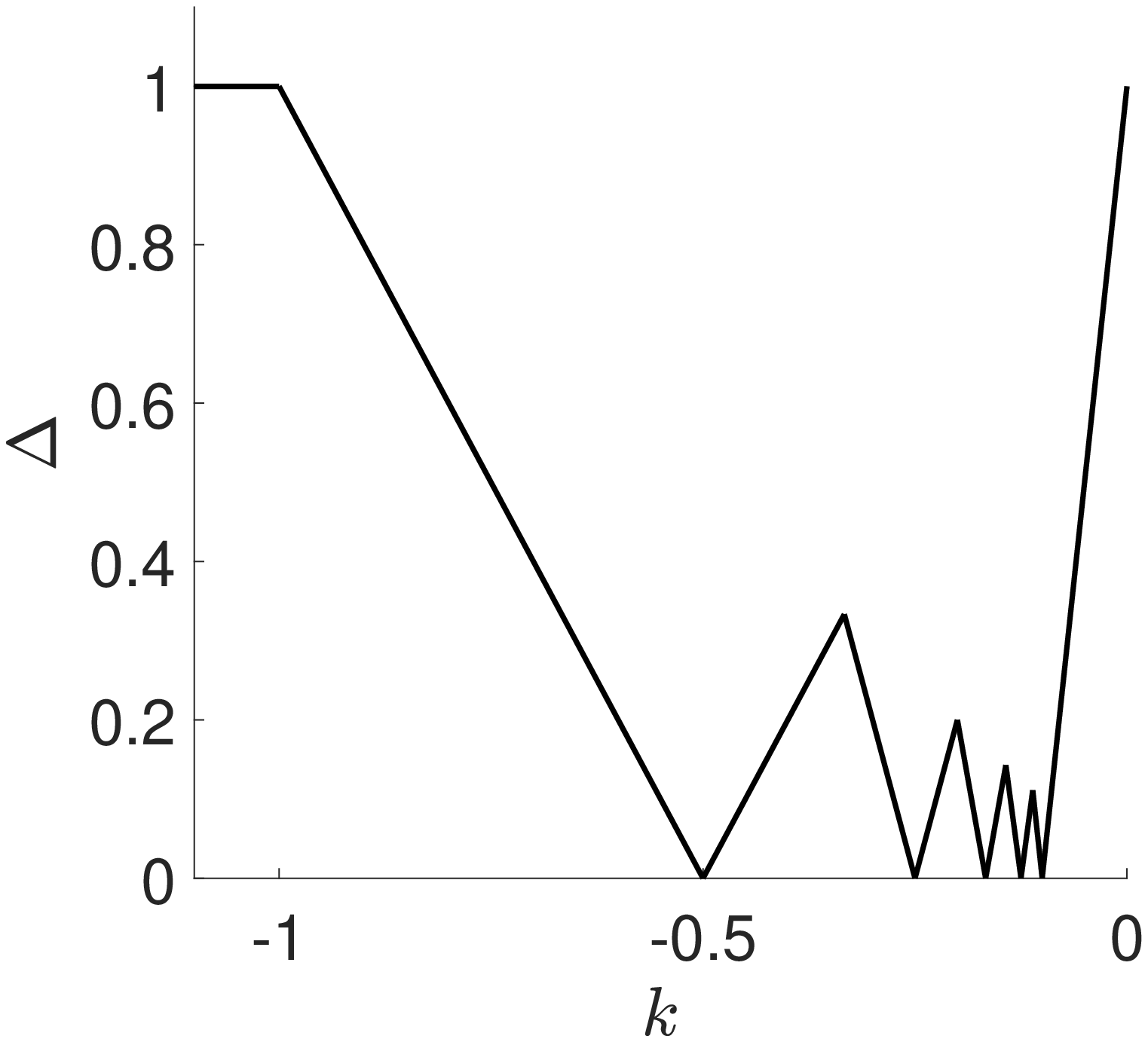}} 
\subfigure[]{\includegraphics[scale=0.27]{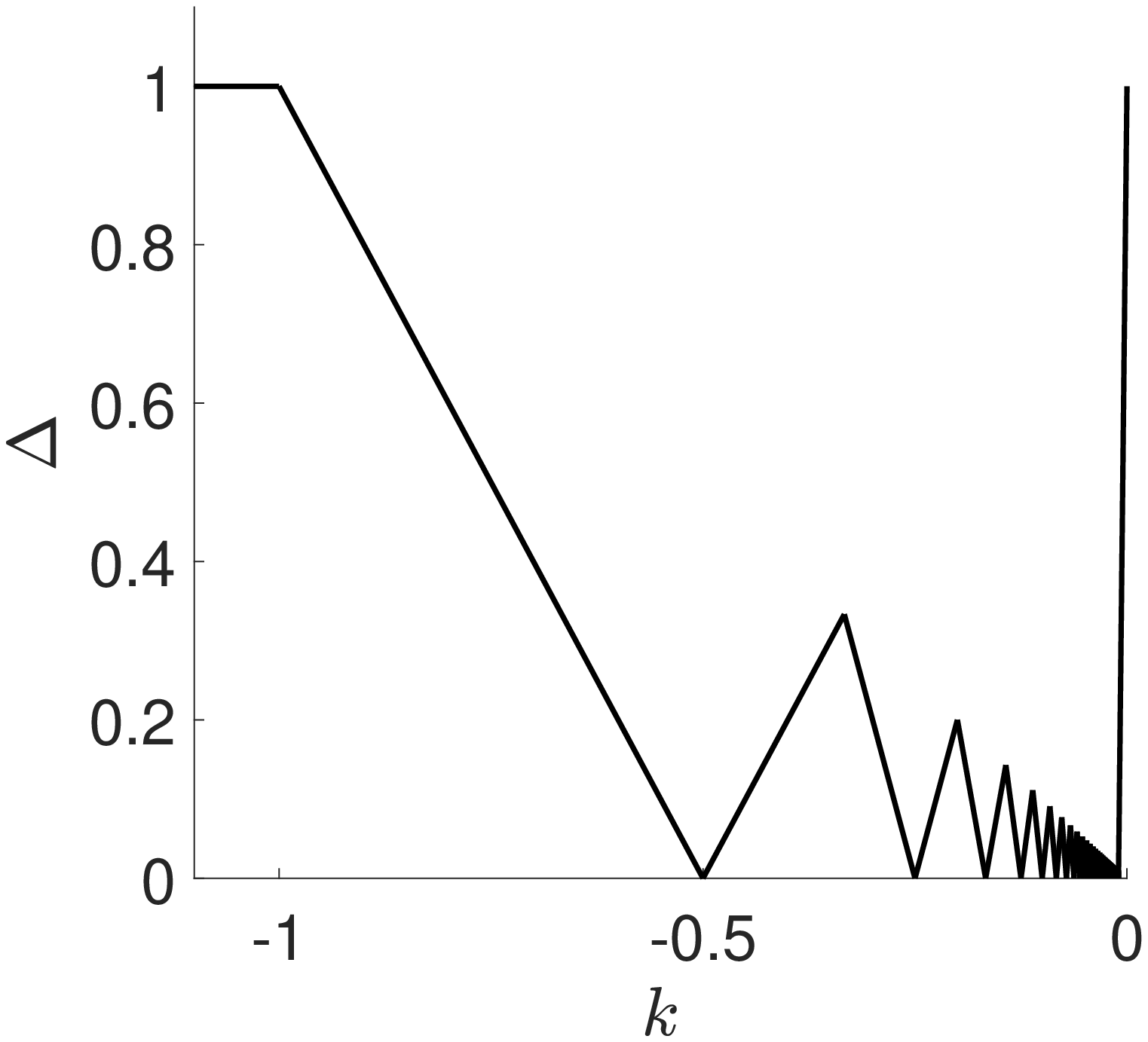}}
}
\begin{caption}
{Ground-state energy gap $\Delta$ as a function of $k$ with $J=1$, and $\Omega=k/2$. (a) $N=5$, (b) $N=11$, (c) $N=101$. }
\end{caption}
\end{figure}

Fig. 2  plots $\Delta$ as a function of $k$ for $N=4,\, 10,\, 100$, and by
contrast, Fig. 3 gives the same plot for odd values $N=5, \,11,\,101$. 
For the odd case, the maximum value of the gap is $J$ for negative values of
$k$, and for the even case the gap is unbounded as $k\rightarrow -\infty$.
It is given by $\Delta= -J-k $ whenever $k\leq -J$.
Furthermore, these figures illustrate that the gap converges to a ``sawtooth'' function,
however the cases of even $N$ and odd $N$ do not converge to the same function.
Indeed, the relationship is one in which the locations of the zeros and peaks of the sawtooth functions are interchanged.
In both instances the convergence is pointwise, which can be proved rigorously. 
In neither case, however, is the convergence uniform with respect to the $||
.||_{\infty}$ norm. This is an explicit example of a number parity effect.

\subsection{Continuum approximation}

Fig. 4 plots $\Delta$ as a function of $\lambda$ for $N=5,\, 11,\, 101$.
It indicates that the gap vanishes in the limit $N\rightarrow\infty$, consistent
with the analysis of Subsection 3.1.  
The gapless regime, which occurs for $k<-1/2$, arises independently of $N$ being
even or odd, but the convergence is not uniform.
In other words, in the continuum approximation, the number parity effect is lost.

\begin{figure}[h]
\center{
\subfigure[]{\includegraphics[scale=0.27]{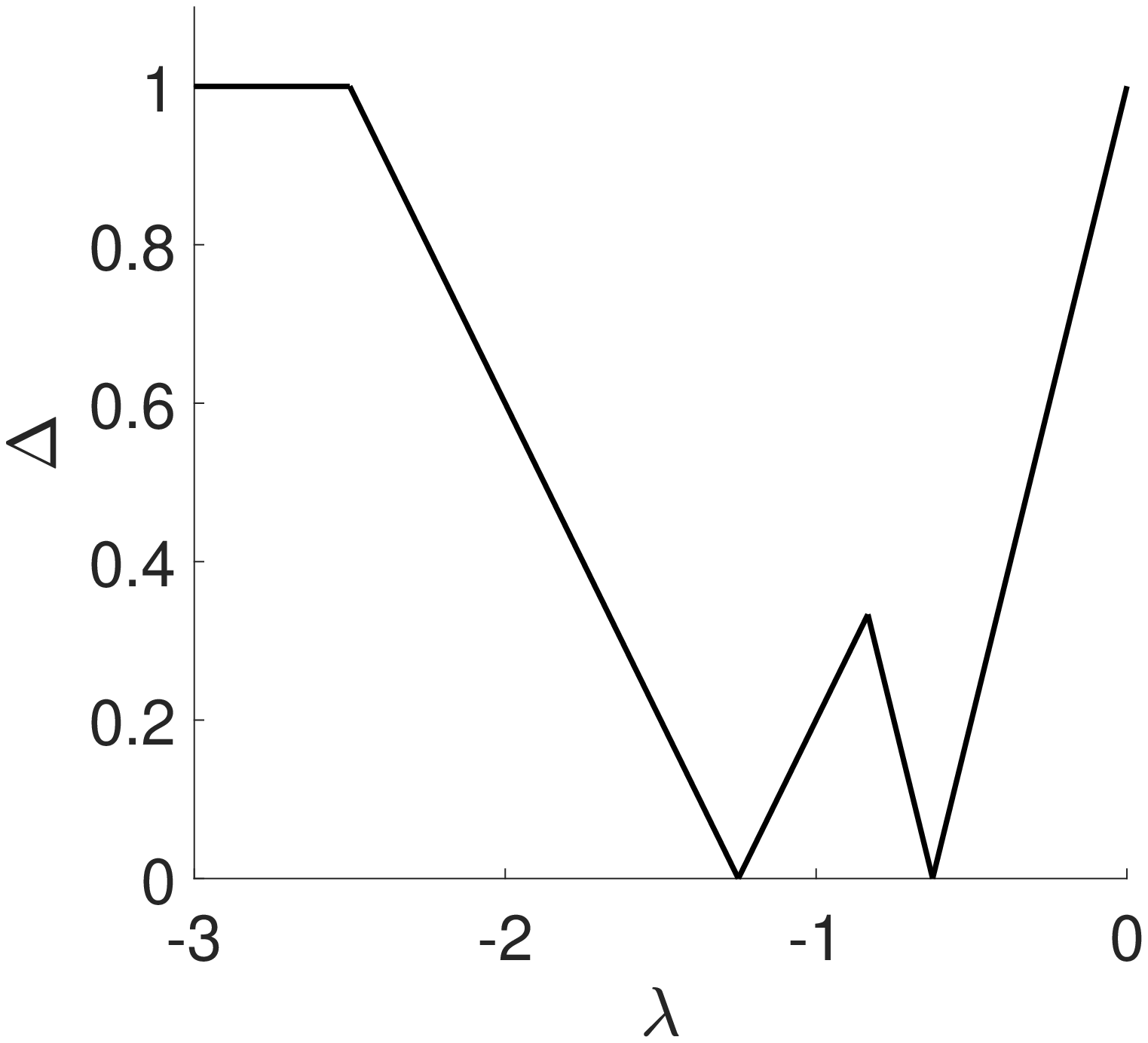}} 
\subfigure[]{\includegraphics[scale=0.27]{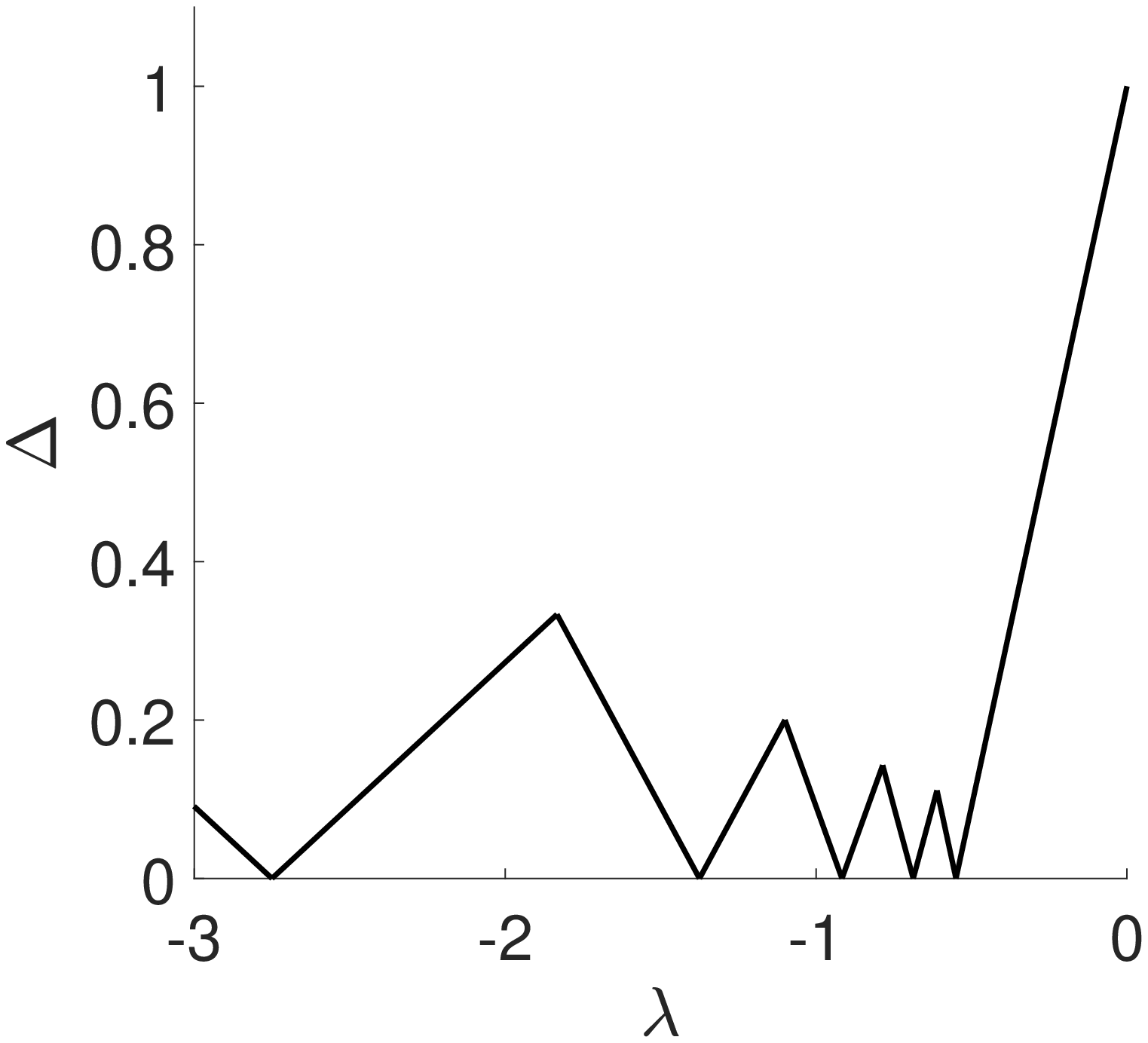}} 
\subfigure[]{\includegraphics[scale=0.27]{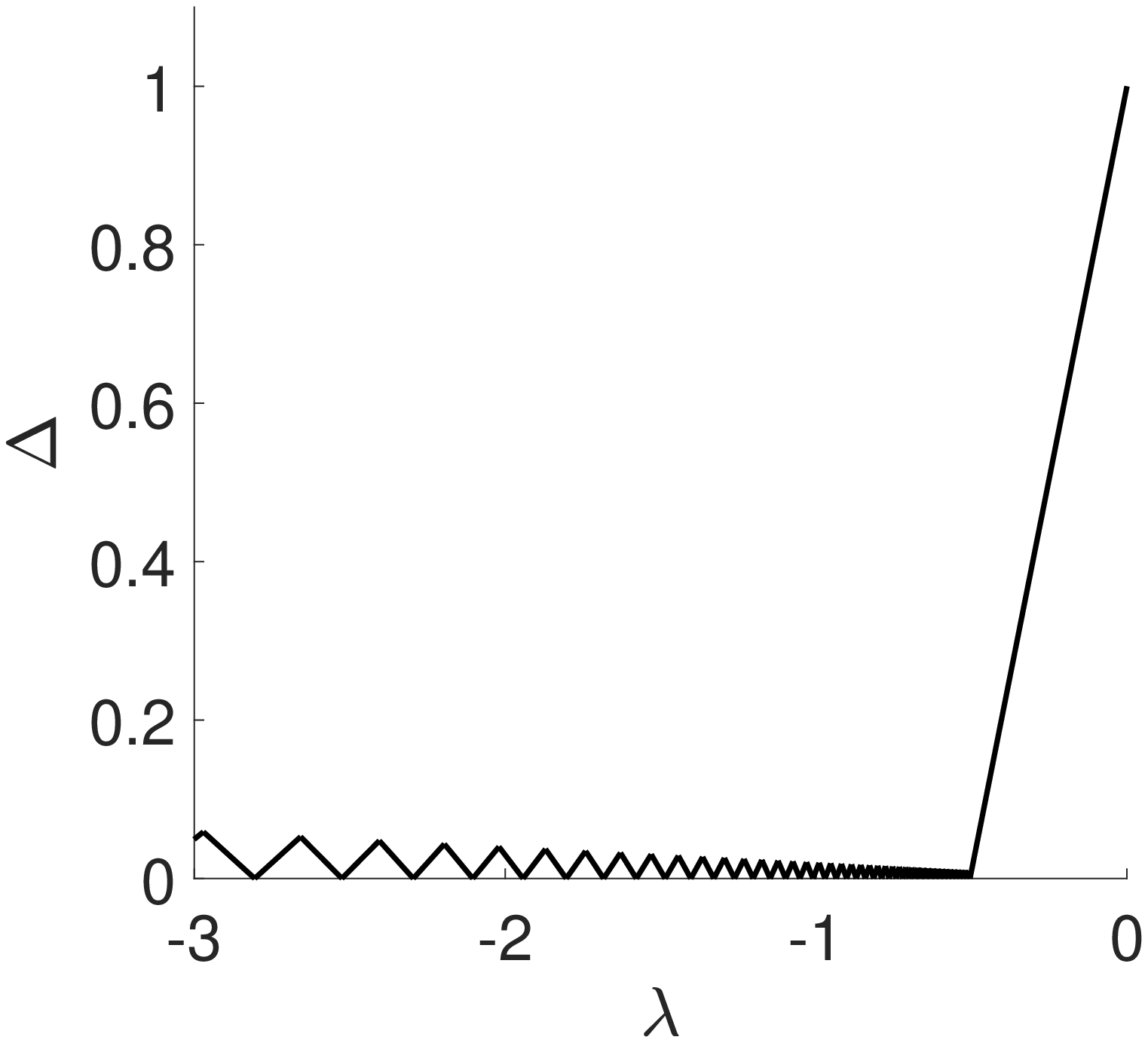}}
}
\begin{caption}
{Ground-state energy gap $\Delta$ as a function of $\lambda$ with $J=1$, and $\Omega=k/2$.  (a) $N=5$, (b) $N=11$, (c) $N=101$. }
\end{caption}
\end{figure}

For other aspects of the system, however, the number parity effect still has a
significant influence. We investigate some further consequences of numberr parity in the remaining sections.






\section{Supersymmetry}

When $N$ is even, $k=J$, and recalling we have fixed $\Omega=k/2$, the Hamiltonian possesses supersymmetry \cite{witten1}. This is expected since at these values we observe multiple two-fold degeneracies and a single state, the ground state, which is non-degenerate, that can be observed in Fig. 1(b). To formalise the result, note that the crossing of energy levels associated with states $|N,q\rangle$ and $|N,q'\rangle$ occurs when (\ref{cross}) holds.
Set $q'=N+1-q$ and define
\begin{align*}
Q&=\sum_{q=1}^{N/2}\sqrt{q q'}|N,q\rangle \langle N,q' |, \\
Q^\dagger&=\sum_{q=1}^{N/2}\sqrt{q q'}|N,q'\rangle \langle N,q |. 
\end{align*}
It is easily verified that
\begin{align}
Q^2=(Q^\dagger)^2=0.
\label{square}
\end{align} 

Define the Hamiltonian
\begin{align}
H=J\left(Q^\dagger Q + Q Q^\dagger - \frac{1}{4}C\right)  
\label{sham}
\end{align}
where $C$ is the $su(2)$ Casimir element. Recall that the eigenvalue of $C$ is $\Lambda= N(N+2)/2$.
It is easy to check using only (\ref{square}) that if $|\Phi\rangle$ is an eigenstate of (\ref{sham}) with eigenvalue $E$ then $Q|\Phi\rangle$ and $Q^\dagger|\Phi\rangle$ are either eigenvectors with the same eigenvalue, or null vectors.
Explicitly from (\ref{sham})
\begin{align}
H&=J\left(-\frac{1}{4}C+\sum_{q=1}^{N/2} q q'\left(|N,q\rangle \langle N,q |+|N,q'\rangle \langle N,q' |\right)\right)  \nonumber \\
&=\frac{J}{8}\left(-N(N+2)|N,0\rangle \langle N,0 |+\sum_{q=1}^{N/2} (8q q'-N(N+2))\left(|N,q\rangle \langle N,q |+|N,q'\rangle \langle N,q' |\right)\right)  \nonumber \\
&=\frac{J}{8}\left(\sum_{q=0}^{N} (8q(N+1-q)-N(N+2))|N,q\rangle \langle N,q |\right)  .
\label{spec}
\end{align}
Setting $k=J$ in (\ref{nrg}) gives the spectrum of (\ref{sham}), as confirmed by (\ref{spec}).   

However, there is no supersymmetry point in the coupling parameter space when  $N$ is odd. One example where this particular parity property has a striking manifestation is in the study of quantum dynamics.


\section{Quantum dynamics}

Let
\begin{align*}
|\Phi\rangle=\frac{1}{\sqrt{N!}}\left(b_1^\dagger\right)^N|0\rangle,
\end{align*}
which represents an initial state such that all particles are in the same site.
Define the expectation value of the fractional atomic imbalance to be 
$$
I={N}^{-1}\langle\Phi|\exp(iHt)(N_1-N_2)\exp(-iHt)|\Phi\rangle, 
$$
where $t$ denotes time.  It can be shown using 
\begin{align*}
\langle N,q|\Phi\rangle= \sqrt\frac{N!}{{2^N q!(N-q)!}}.
\end{align*}
that a simple expression for $I$ is obtained:
\begin{align}
I
&=\cos(Jt)(\cos(kt))^{N-1}.
\label{I}
\end{align}

At the supersymmetric point $k=J$ when $N$ is even, it is apparent that $0\leq  I\leq 1$. For odd $N$ at the same value of coupling parameters it is apparent that $-1\leq I \leq 1$. Thus the even $N$ case exhibits a type of self-trapping behaviour, while the odd $N$ does not. While the phenomenon of self-trapping is well-known \cite{Mil97,Leg01}, such a parity influence on self-trapping does not appear to have been previously identified.

\begin{figure}[t] 
{\includegraphics[scale=0.24]{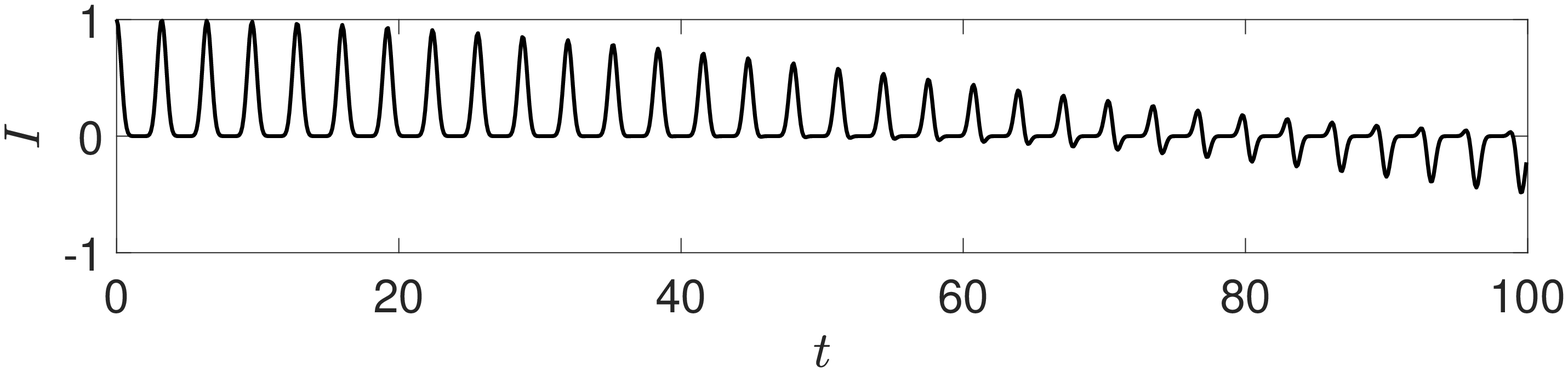}}  
{\includegraphics[scale=0.24]{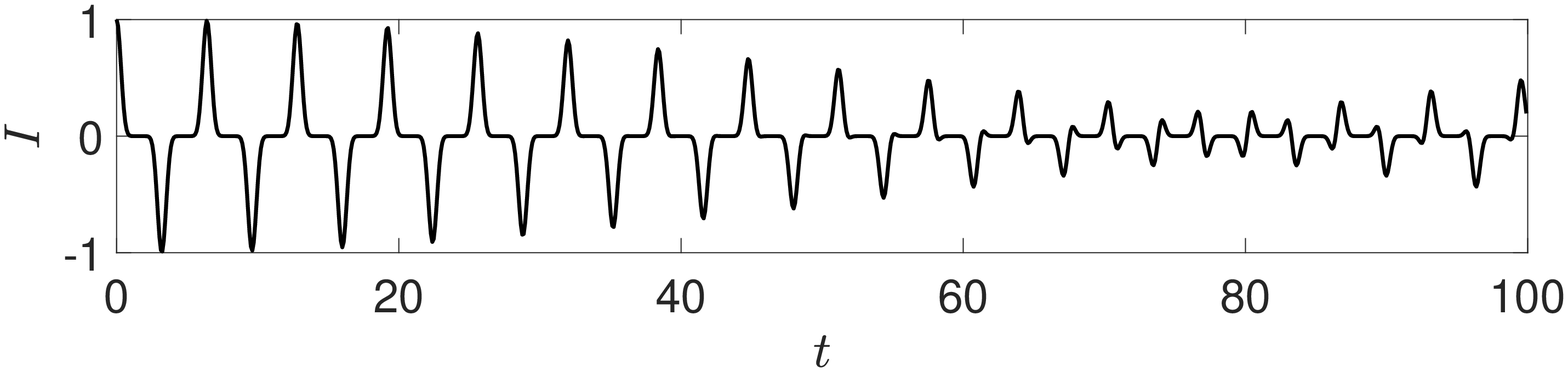}}  \\
{\includegraphics[scale=0.24]{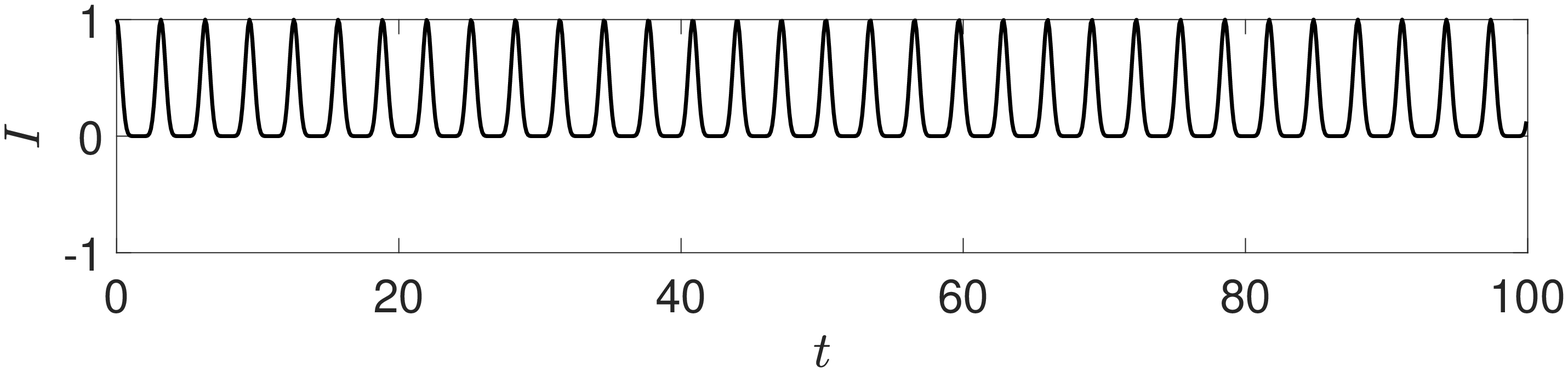}}  
{\includegraphics[scale=0.24]{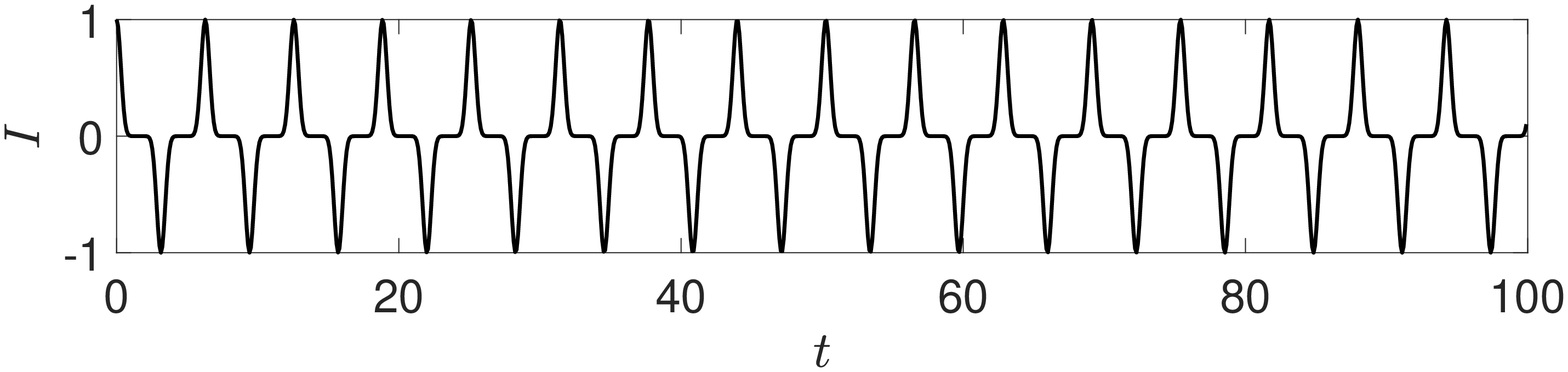}} \\
{\includegraphics[scale=0.24]{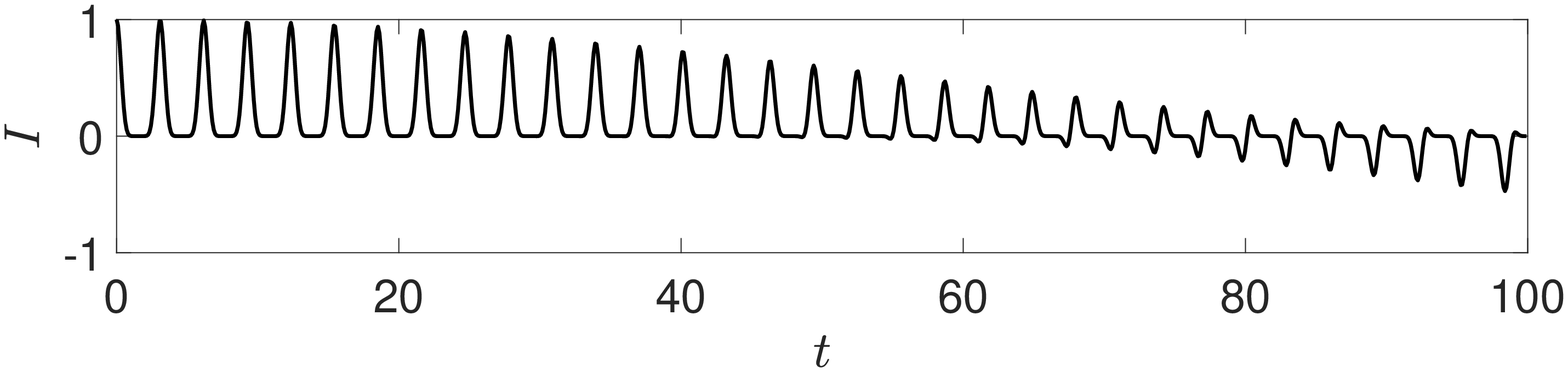}} 
{\includegraphics[scale=0.24]{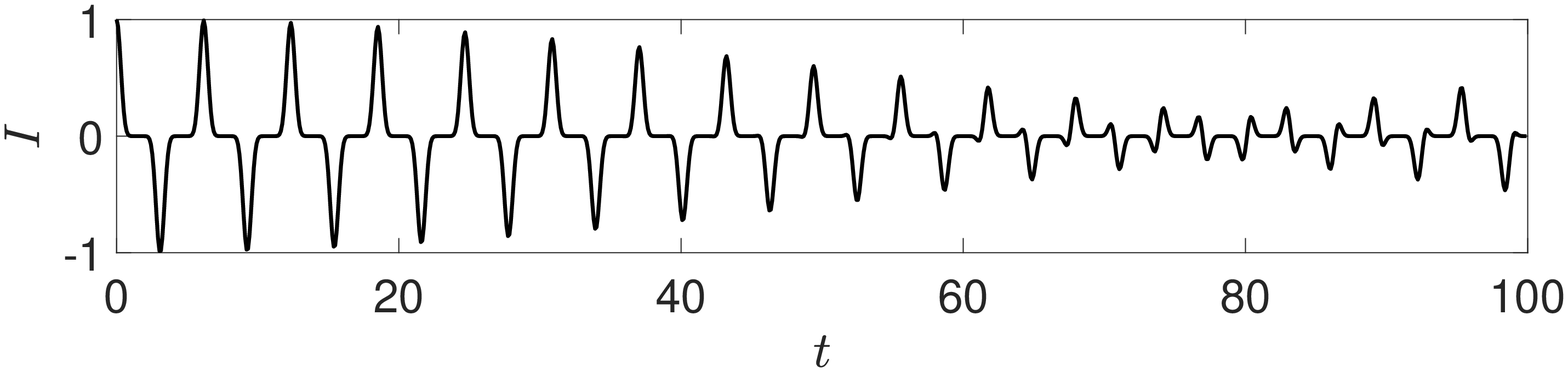}}  
\begin{caption}
{Expectation value of the fractional atomic imbalance $I$ as a function of $t$ for $J=1$ and $\Omega=k/2$. On the left, $N=10$, on the right, 
 $N=11$. From top to bottom $k=\pm 49/50,\, \pm 1,\, \pm 51/50$.}
\end{caption}
\label{fig:minusone}
\end{figure}

\begin{figure}[h!] 
{\includegraphics[scale=0.24]{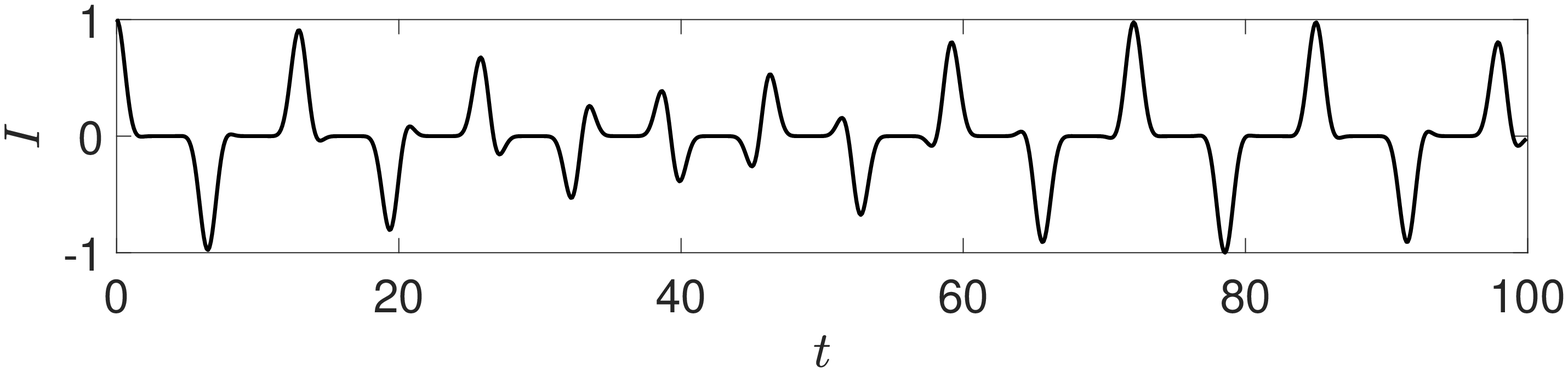}}  
{\includegraphics[scale=0.24]{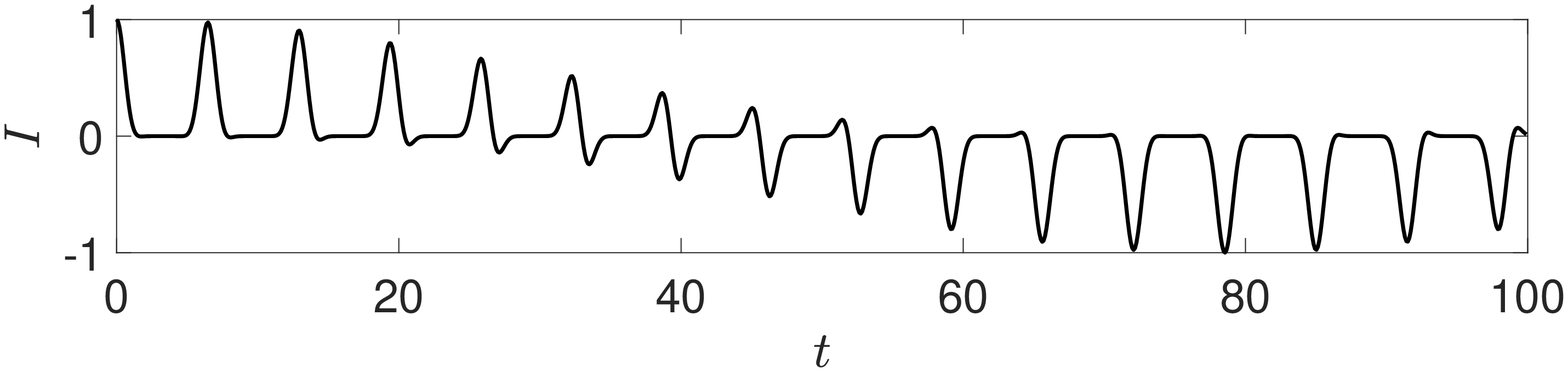}}  \\
{\includegraphics[scale=0.24]{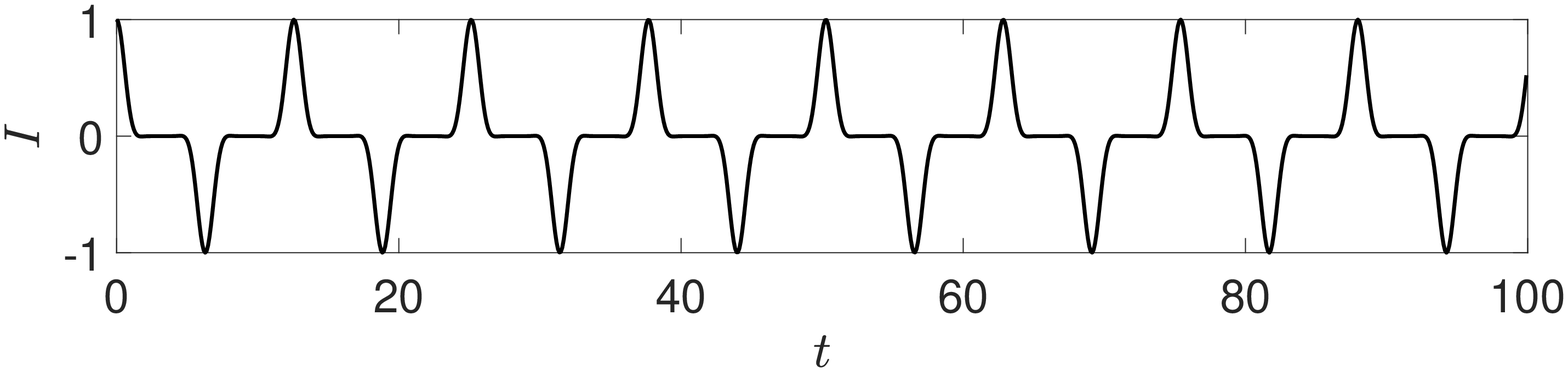}}  
{\includegraphics[scale=0.24]{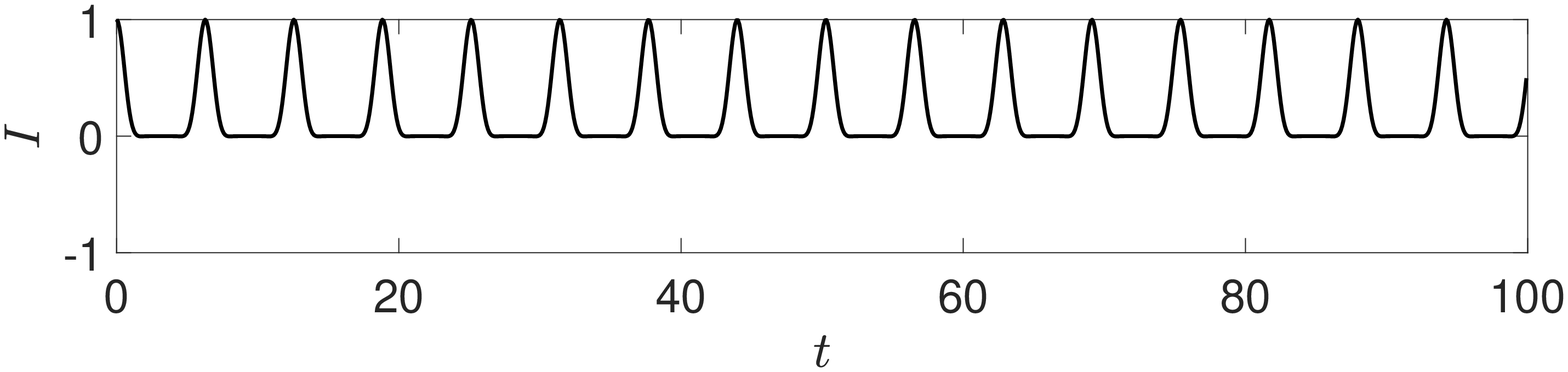}} \\
{\includegraphics[scale=0.24]{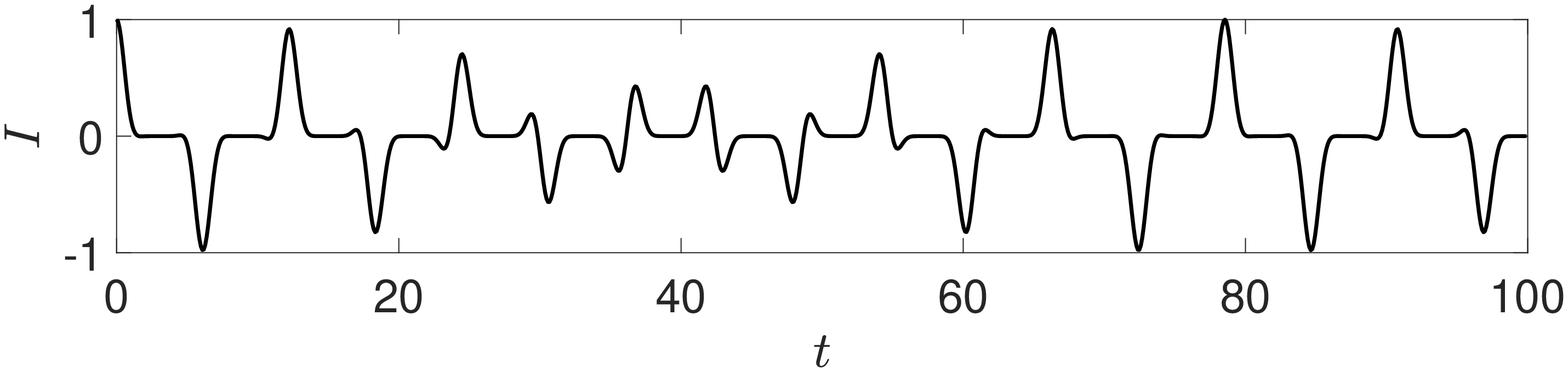}} 
{\includegraphics[scale=0.24]{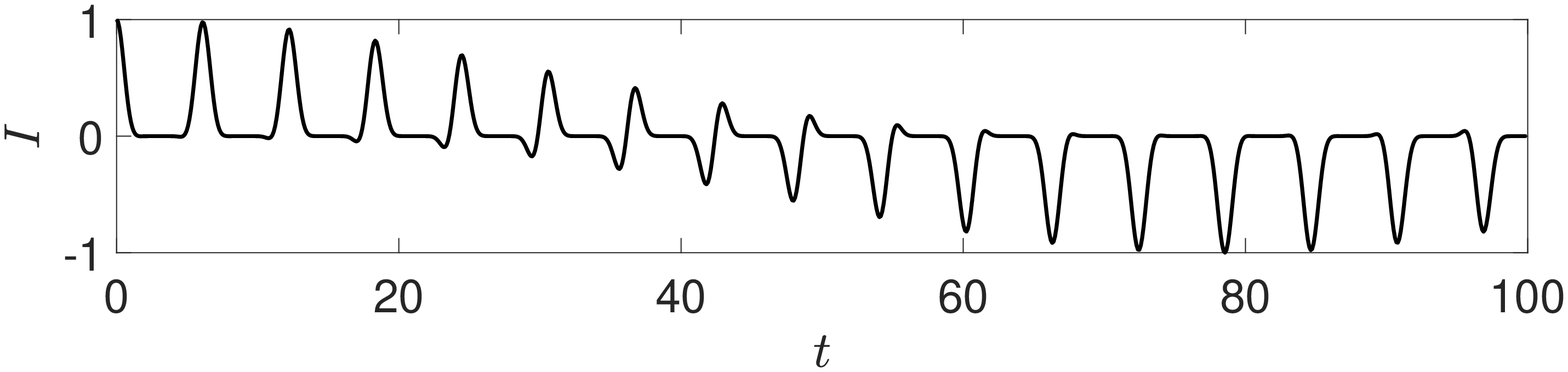}}  
\begin{caption}
{Expectation value of the fractional atomic imbalance $I$ as a function of $t$ for $J=1$ and $\Omega=k/2$. On the left, $N=10$, on the right, 
 $N=11$. From top to bottom $k =\pm 12/25,\, \pm 1/2,\, \pm 13/25$.}
\end{caption}
\label{fig:minushalf}
\end{figure}

Since the expression (\ref{I}) is an even function of $k$, exactly the same dynamical behaviour occurs for $k=-J$.   For even $N$ this corresponds to the smallest value of $k$ such that the ground-state energy gap is zero. In contrast, for odd $N$  the smallest value of $k$ for which the ground-state energy gap is zero is $k=-J/2$. Illustrative examples of the expectation values for the  fractional atomic imbalance at these parameter values are provided in Figs. 5 and 6 for $N=10$ and $N=11$.    It is clear that the number-parity significantly influences the character of the dynamical behaviour. This remains true for different parameter vales, although the 
effects are not so pronounced. An example is given in Fig. 7, with parameter values in the vicinity of $k=-1/N$ for $J=1$. 

\begin{figure}[h!]
{\includegraphics[scale=0.24]{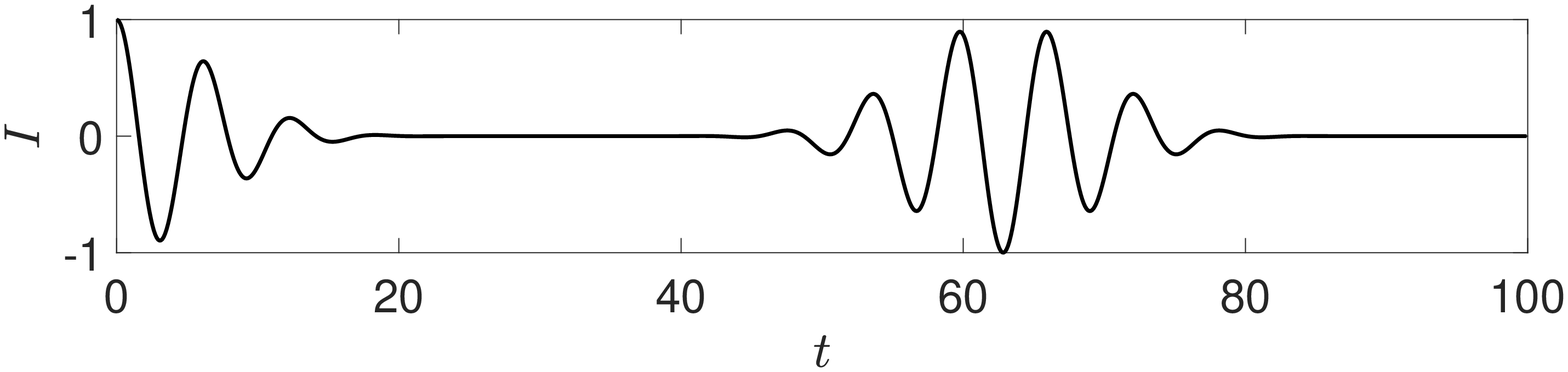}}  
{\includegraphics[scale=0.24]{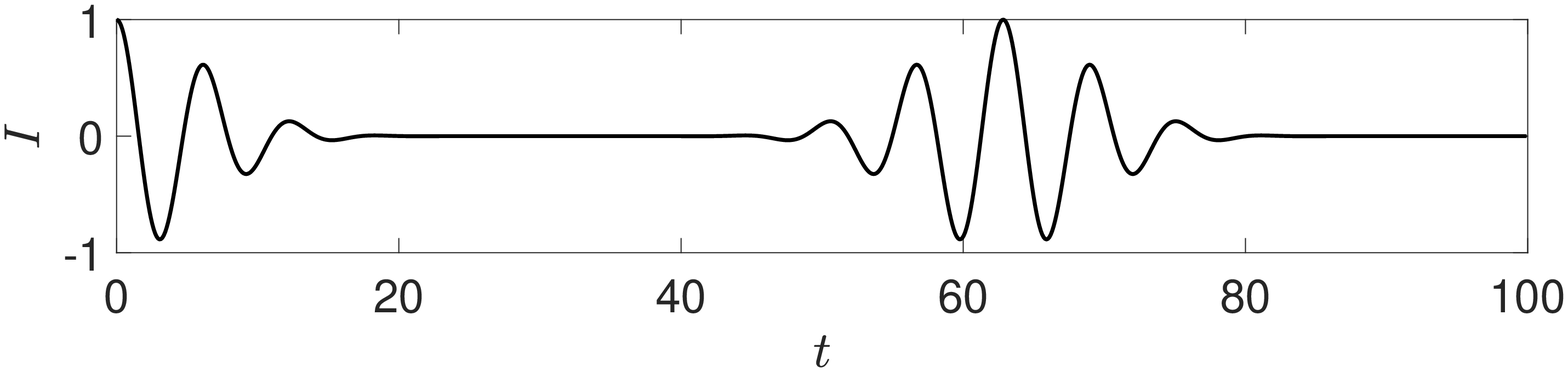}}  \\
{\includegraphics[scale=0.24]{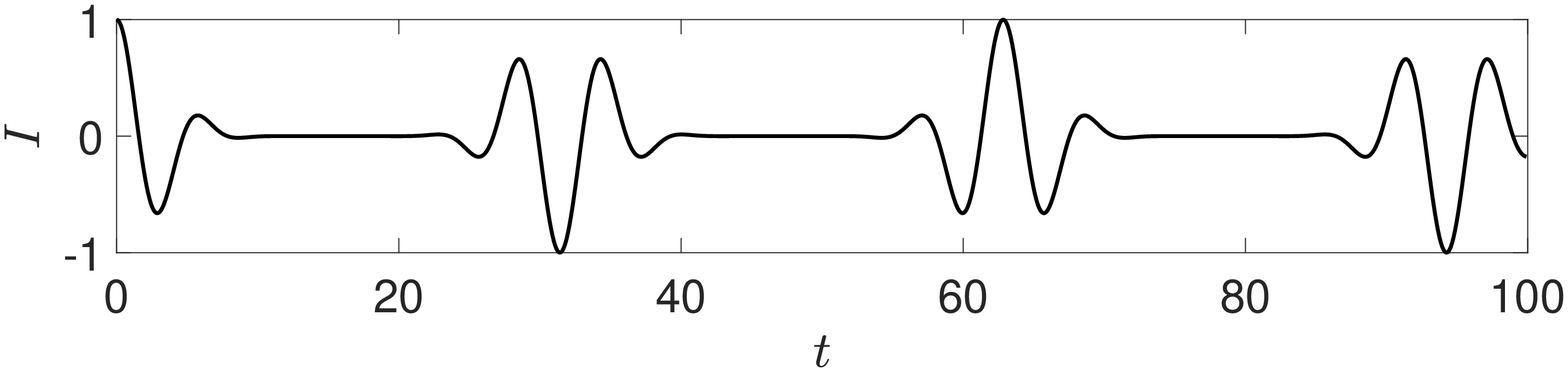}}  
{\includegraphics[scale=0.24]{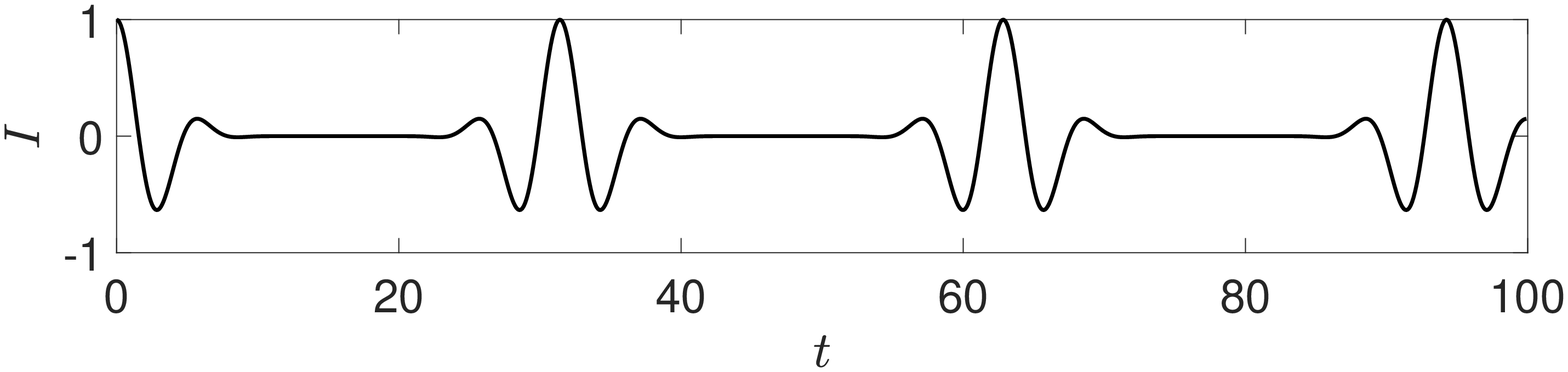}} \\
{\includegraphics[scale=0.24]{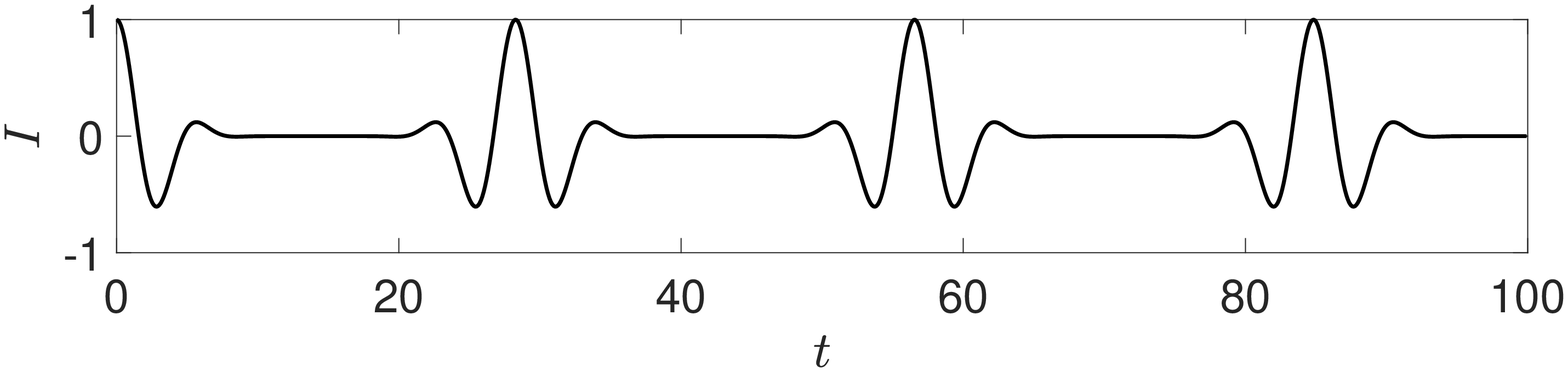}} 
{\includegraphics[scale=0.24]{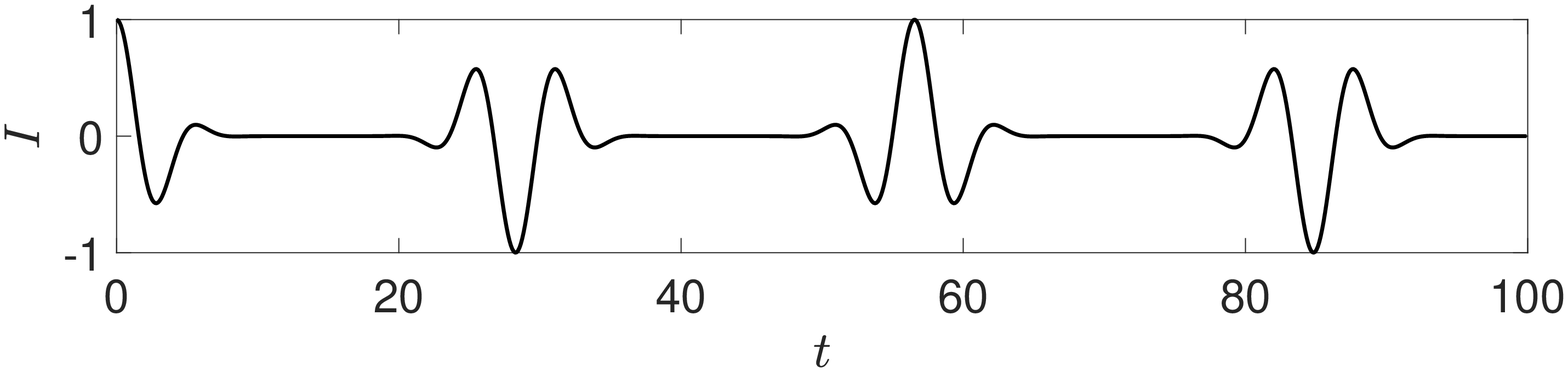}}  
\begin{caption}
{Expectation value of the fractional atomic imbalance $I$ as a function of $t$ for $J=1$ and $\Omega=k/2$. On the left, $N=10$, on the right, 
 $N=11$. From top to bottom $k=\pm 1/20,\, \pm 1/10,\, \pm 1/9$.}
\end{caption}
\end{figure}

\section{Conclusion}  
 
We have studied an extension of the familiar two-site Bose-Hubbard model that includes a second-order tunneling term. This model is known to exhibit three phases determined by fixed-point bifurcations, and in the present work a detailed analysis has been undertaken along a line of the coupling parameter space that includes the boundary between phase-locking and self-trapping phases. All energy levels on this line can be computed analytically, and from this result it was identified that significant number-parity effects are present. In particular, the influence of number-parity on the ground-state energy gap, and the dynamics of the fractional atomic imbalance, were investigated. 

Mathematically, the model considered here is equivalent to the Lipkin-Meshkov-Glick (LMG) model of nuclear physics, which can be seen through the spin representation (\ref{lmg}). The LMG model been studied through an exact Bethe Ansatz solution \cite{ld13}, although the exact solution has a different form to that of \cite{rlif17}. In \cite{ld13} the analysis was conducted using a choice of coupling parameters such that the region of level crossing is treated as a gapless region in the limit of large particle number. Our results indicate that there may be new insights to be gained for the LMG model by choosing a different form of coupling parameters. Such an approach has recently been applied to the attractive one-dimensional Bose gas \cite{pc16}, whereby a distinction is made between the zero density thermodynamic limit and the weakly interacting thermodynamic limit which are obtained by different scaling of parameters as the system size increases.    

\ack This research was supported by the Australian Research Council through Discovery Project DP150101294.   
We thank the mathematical research institute MATRIX in Creswick, Australia, where part of the work was undertaken during the programs {\it Integrability in low-dimensional quantum systems}, and {\it Combinatorics, statistical mechanics, and conformal field theory}.

\end{document}